\documentclass{iaus}
\usepackage{graphicx}

\title[The age of IZw18] 
{IZw18, or the picture of Dorian Gray: the more you watch it, the older it gets}

\author[Tosi et al.]   
{M.~Tosi$^1$,%
 A.~Aloisi$^2$, J.~Mack$^3$, \and M.~Maio$^1$ }

\affiliation{$^1$INAF - Osservatorio Astronomico di Bologna,
Bologna, Italy \break email:monica.tosi@oabo.inaf.it\\[\affilskip]
$^2$ESA and Space Telescope Science Institute, Baltimore, USA
\\[\affilskip] $^3$Space Telescope Science Institute, Baltimore, USA 
}

\pubyear{2006}
\volume{235}  
\pagerange{???}
\date{?? and in revised form ??}
\jname{Galaxy Evolution across the Hubble Time}
\editors{F.Combes \& J. Palous, eds.}
\begin{document}

\maketitle

\begin{abstract}
IZw18 has been recurrently claimed to be a young galaxy, but 
stars of increasingly older ages are found every time deeper magnitude 
levels are reached with high-resolution photometry: from the original few Myrs  
to, possibly, several Gyrs. 
We summarize the history of IZw18's age and an HST project which will allow us 
to derive both its distance and age. 
\end{abstract}

\firstsection 
\section{Introduction}
The Blue Compact Dwarf galaxy IZw18 is the most metal poor star-forming
galaxy ever discovered, with a metallicity between 1/30 and 1/50 of solar, 
depending on the adopted solar value 
(e.g., \cite[Searle \& Sargent 1972]{}, \cite[Izotov \& Thuan 2004]{}). 
IZw18 is not only extremely metal poor, but also very 
blue and full of gas, three properties that originally suggested it to be 
forming now its first stars and make it the system in the local Universe most 
similar to primeval galaxies. 
Soon after its discovery, the basic question became whether IZw18 is so metal 
poor because a) it has started so recently to form stars that they haven't had 
time to significantly pollute the medium, or b) because its star formation (SF) 
activity, although occurring over a long period of time, has proceeded at a 
rate too low for an 
efficient chemical enrichment, or c) because strong galactic winds removing from 
the system most of the metals have accompanied a {\it normal} SF activity. 
More than thirty years later, its age and distance are still subjects of hot 
debates. 

The safest way to estimate the age of nearby galaxies is from the 
Colour-Magnitude Diagram (CMD) of their resolved stellar populations, but for
IZw18, at 10--20 Mpc, this is feasible only with HST and yet
challenging. Our novel starts with the advent of HST.

\section{IZw18 looks older everytime we watch it}

\begin{figure}
\centerline{
\scalebox{0.40}{%
\includegraphics{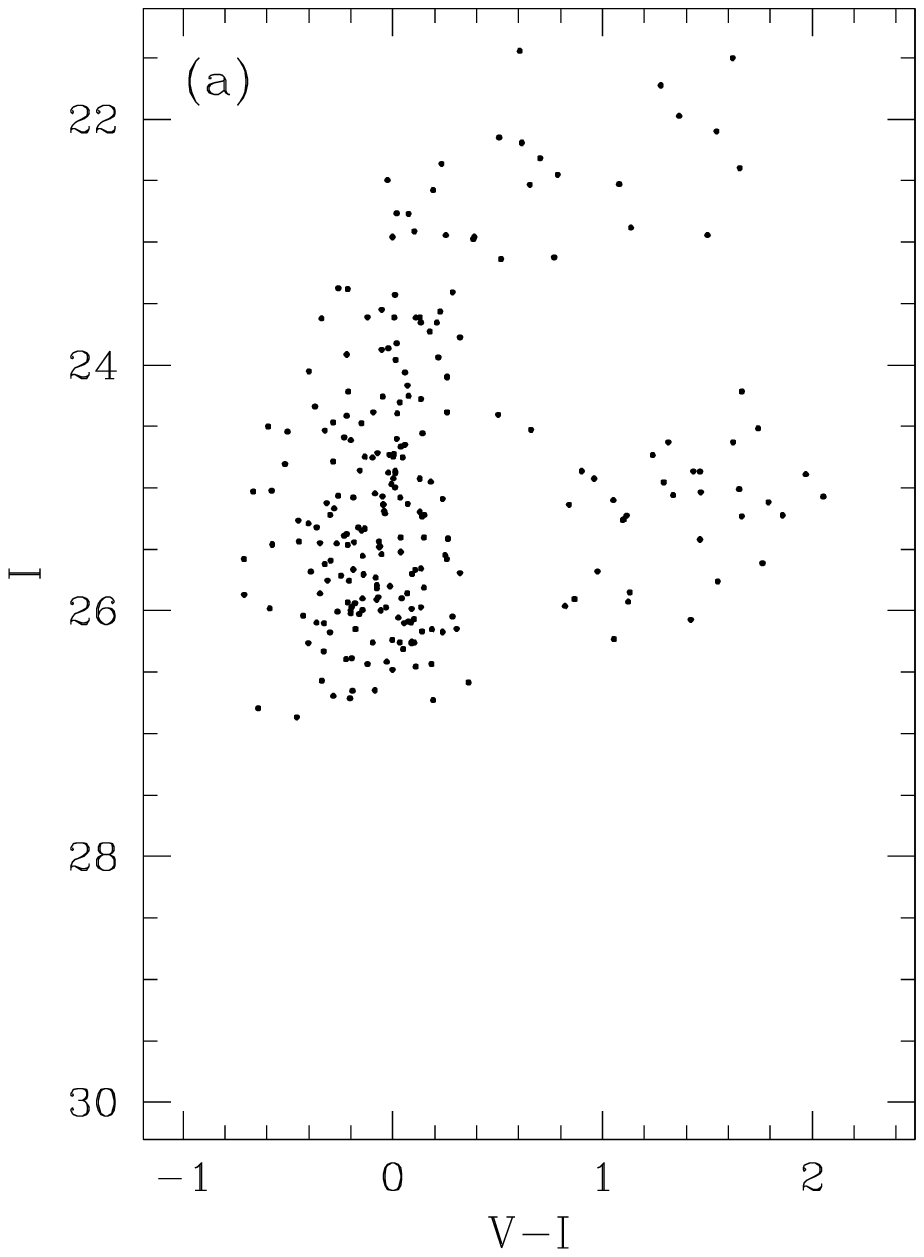}%
}
\scalebox{0.40}{%
\includegraphics{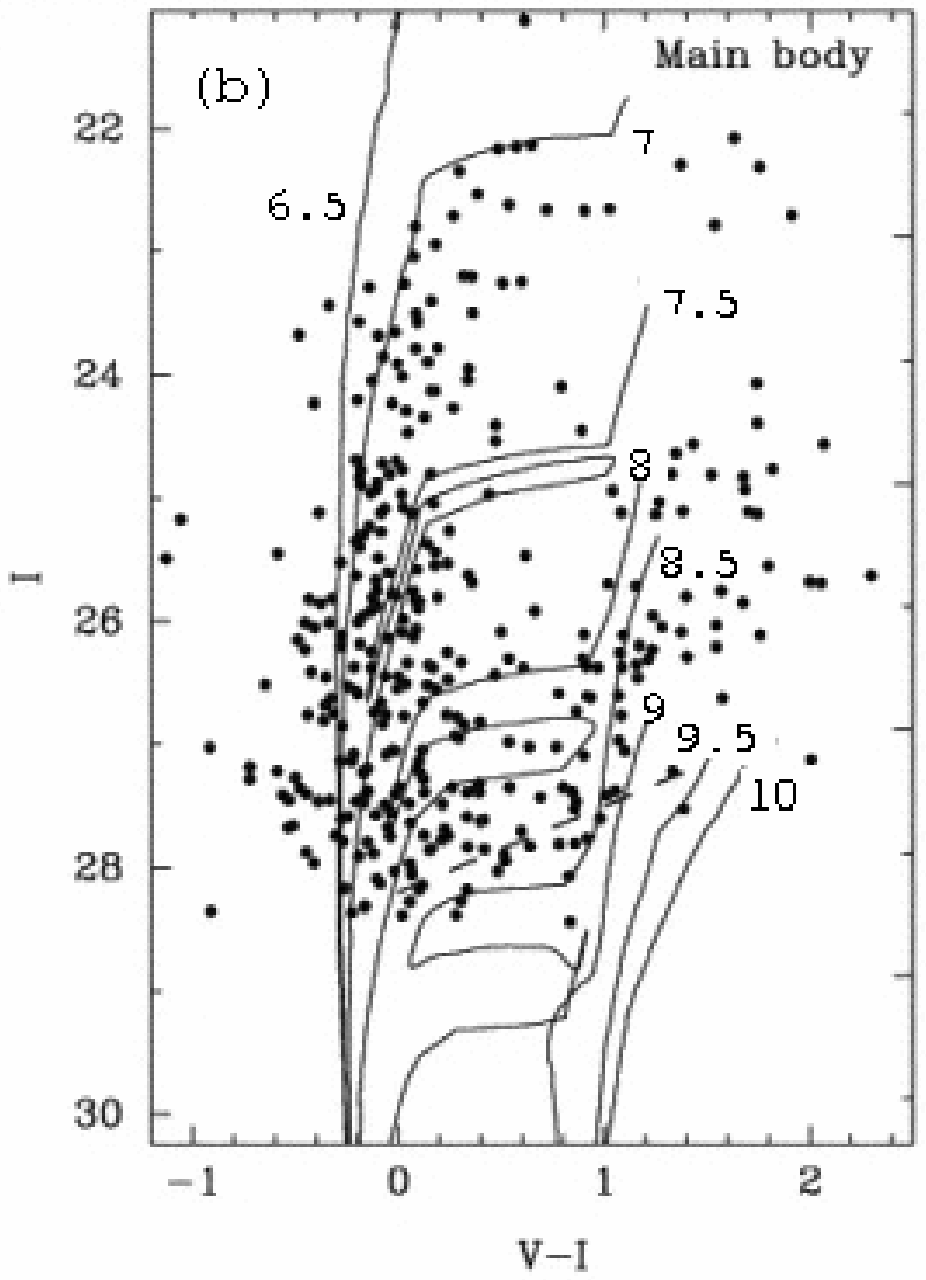}%
}
\scalebox{0.40}{%
\includegraphics{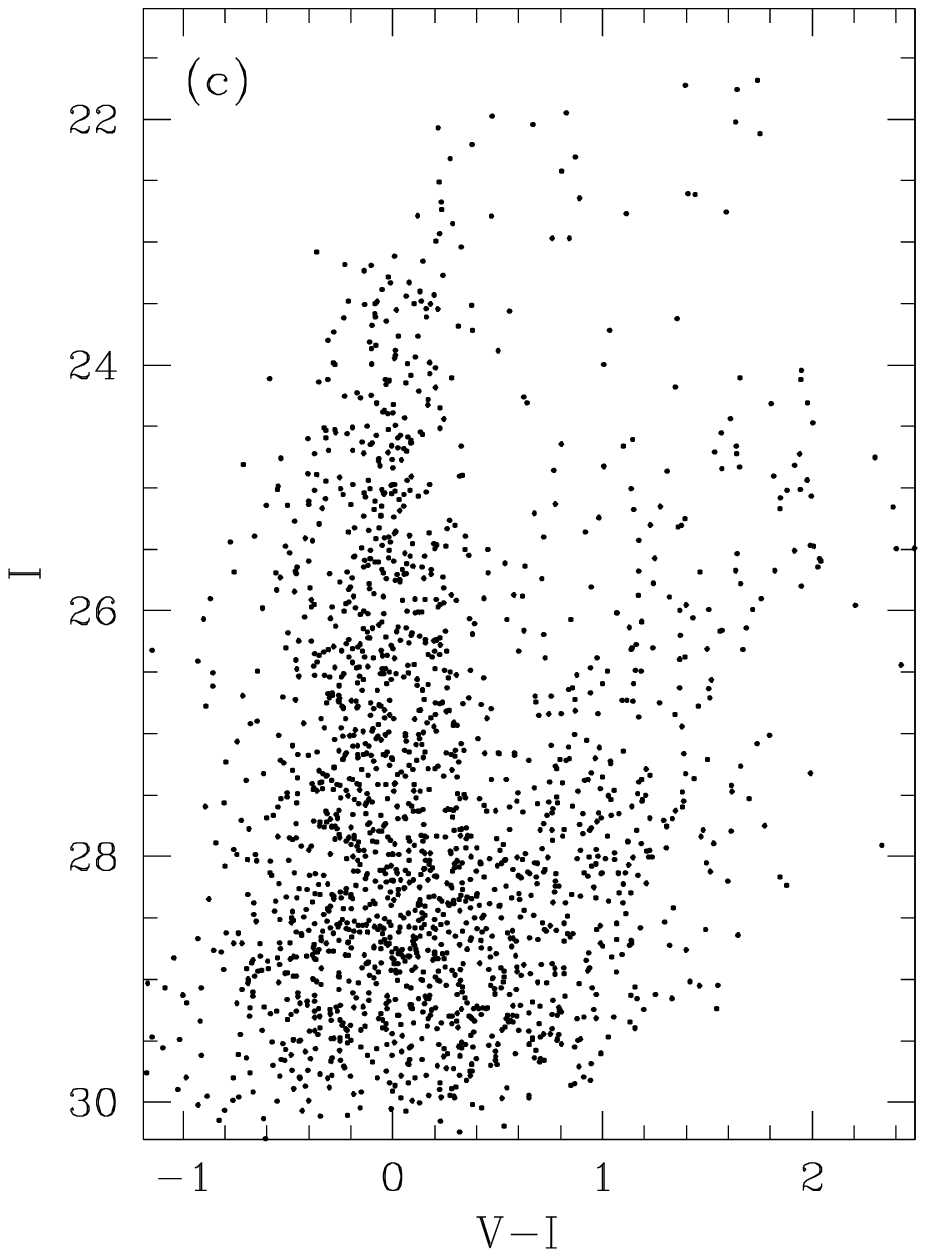}%
} 
}
\caption{CMDs of the stars resolved in the Main Body of IZw18: 
    (\textit{a}) by \cite{Al99} from WFPC2, 
    (\textit{b}) by IT04 from ACS,
    (\textit{c}) by us from the same ACS images as IT04.}\label{fig:cmd}
\end{figure}

The first CMDs derived from HST-WFPC2 images 
(\cite[Hunter \& Thronson 1995]{HT95}) 
seemed to confirm that the stars in IZw18 are only a few 10 Myr old, until  
\cite[Aloisi, Tosi \& Greggio (1999)]{Al99} identified a number of faint red 
stars falling in the Asymptotic Giant Branch (AGB) region of the CMD, if not 
on the RGB (Fig.~\ref{fig:cmd}\,(\textit{a})). A result implying that
stars at least 500 Myr old are present in the galaxy, and 
confirmed by  \cite{O00}, who independently measured the same AGB stars on 
near-infrared HST-Nicmos images. 
When the ACS became available on HST, with much higher sensitivity, 
\cite[Izotov \& Thuan (2004, hereafter IT04)]{IT04} 
reobserved IZw18 and confirmed the existence of the AGB, but claimed that 
no star on the RGB could be detected, thus putting an
upper limit of 500 Myr to the age of the system. However, when other people 
have reduced the same ACS images, they have reached much fainter magnitudes 
than IT04. The CMD derived by IT04 from their HST-ACS images is shown in 
Fig.~\ref{fig:cmd}\,(\textit{b}) together with theoretical isochrones of
the labelled log(age) and assuming 
a distance of 15 Mpc. The CMD obtained by us from the same data is 
shown in Fig.~\ref{fig:cmd}\,(\textit{c}). \cite{M05} also obtained a CMD
like ours, more than two mags deeper than IT04's. 
It is striking that IT04's photometry stops just 
where the tip of the RGB is expected (I $\simeq 27.2$ at 15 Mpc), 
as indicated by the two redder isochrones. We believe that the increase in 
number of stars occurring in our CMD just at I $\simeq 27.2$ despite the 
increasing level of incompleteness is strongly 
suggestive of the presence of the RGB. As shown by the isochrone labels in 
Fig.~\ref{fig:cmd}\,(\textit{b}), this would imply an age significantly 
older than 2 Gyr, possibly 10 Gyr or more. However, this indication 
cannot be confirmed until better estimates of the galaxy distance are 
available to precisely constrain the expected magnitude of the RGB tip.

This is why we have re-observed IZw18 for 24 HST-ACS orbits at 12 different epochs
(prop. 10586, PI Aloisi) to perform time
series photometry, identify Cepheids, measure their light curves and infer 
their distance from the period-luminosity and period-luminosity-colour 
relations, following the method applied in the HST H$_0$ key-project 
(\cite[Saha et al.\ 2006]{S06} and references therein). 
To this aim we are also computing non-linear convective pulsation models for 
the metallicity of IZw18, which will be used for safer light 
curve analyses and distance derivation, as done
for RR Lyraes in the LMC by \cite{MC05}. The combination of
this new photometry with the archival data will allow us to derive 
deeper CMDs to better search for RGB stars. Independent distance estimates 
will be inferred from the brightness of the Carbon stars and of the RGB tip 
(if any).
We have already identified $\sim$30 variables and the data analysis is 
in progress: {\it stay tuned on the IZw18 age serial !}


\begin{thebibliography}{}

\bibitem[Aloisi \etal\ (1999)]{Al99}
   {Aloisi, A., Tosi, M., Greggio, L.} 1999,      \textit{AJ} 118, 302
   
\bibitem[Hunter \& Thronson (1995)]{HT95} 
   {Hunter, D.A. \& Thronson, H.A. } 1995,      \textit{ApJ} 452, 238

\bibitem[Izotov \& Thuan (2004)]{IT04}
  {Izotov, Y.I. \& Thuan, T.X.} 2004,      \textit{ApJ} 616, 768

\bibitem[Marconi \& Clementini (2005)]{MC05}
   {Marconi, M. \& Clementini, G.}  2005,    \textit{AJ} 129, 2257

\bibitem[Momany et al. (2005)]{M05}
   {Momany, Y., Held, E.V., Saviane, I., Bedin, L.R., Gullieuszik, M., 
   Clemens, M., Rizzi, L., Rich, M.R., Kuijken, K.} 2005,     
   \textit{A\&A} 439, 111
 
\bibitem[Ostlin (2000)]{O00}
  {Ostlin, G.}  2000,    \textit{ApJ} (Letters) 535, L99
  
\bibitem[Saha \etal\ (2006)]{S06}
  {Saha, A., Thim, F., Tammann, G. A., Reindl, B., Sandage, A.} 2006, 
   \textit{ApJS} 165, 108

\bibitem[Searle \& Sargent (1972)]{SS72}
  {Searle, L. \& Sargent, W.L.W.}  1972,    \textit{ApJ} 173,25

   
\end{thebibliography}
\end{document}